\documentclass[final,5p,times,twocolumn]{elsarticle}

\usepackage{amsmath}
\usepackage{amssymb}
\usepackage{graphicx}
\usepackage{color}
\usepackage{hyperref}

\usepackage[mathscr,scaled=1.15]{urwchancal}
\DeclareFontFamily{OT1}{pzc}{}
\DeclareFontShape{OT1}{pzc}{m}{it}%
{<-> s * [1.15] pzcmi7t}{}
\DeclareMathAlphabet{\mathpzc}{OT1}{pzc}{m}{it}

\definecolor{purple}{rgb}{0.5,0,0.5}
\definecolor{blue}{rgb}{0.0,0,0.9}

\biboptions{sort&compress}

\journal{Physics Letters B}

\begin{document}

\begin{frontmatter}

\title{Basic features of the pion valence-quark distribution function}

\author[UA]{Lei Chang}
\author[CEA]{C\'edric Mezrag}
\author[CEA]{Herv\'e Moutarde}
\author[ANL]{Craig D. Roberts}
\author[Huelva]{Jose Rodr\'iguez-Quintero}
\author[KSU]{Peter C. Tandy}

\address[UA]{CSSM, School of Chemistry and Physics
University of Adelaide, Adelaide SA 5005, Australia}
\address[CEA]{Centre de Saclay, IRFU/Service de Physique Nucl\'eaire, F-91191 Gif-sur-Yvette, France}
\address[ANL]{Physics Division, Argonne National Laboratory, Argonne, Illinois 60439, USA}
\address[Huelva]{Departamento de F\'isica Aplicada, Facultad de Ciencias Experimentales, Universidad de Huelva, Huelva E-21071, Spain}
\address[KSU]{Center for Nuclear Research, Department of
Physics, Kent State University, Kent, Ohio 44242, USA}

\date{18 June 2014}

\begin{abstract}
$\,$\\[-7ex]\hspace*{\fill}{\emph{Preprint no}. ADP-14-20/T878}\\[1ex]
The impulse-approximation expression used hitherto to define the pion's valence-quark distribution function is flawed because it omits contributions from the gluons which bind quarks into the pion.  A corrected leading-order expression produces the model-independent result that quarks dressed via the rainbow-ladder truncation, or any practical analogue, carry all the pion's light-front momentum at a characteristic hadronic scale.  Corrections to the leading contribution may be divided into two classes, responsible for shifting dressed-quark momentum into glue and sea-quarks.  Working with available empirical information, we use an algebraic model to express the principal impact of both classes of corrections.  This enables a realistic comparison with experiment that allows us to highlight the basic features of the pion's measurable valence-quark distribution, $q^\pi(x)$; namely, at a characteristic hadronic scale, $q^\pi(x) \sim (1-x)^2$ for $x\gtrsim 0.85$; and the valence-quarks carry approximately two-thirds of the pion's light-front momentum.
%
\end{abstract}

\begin{keyword}
deep inelastic scattering \sep
Drell-Yan process \sep
dynamical chiral symmetry breaking \sep
Dyson-Schwinger equations \sep
$\pi$-meson \sep
parton distribution functions

\smallskip



\end{keyword}

\end{frontmatter}

\noindent\textbf{1.$\;$Introduction}.
With the advent of the constituent-quark model, the pion came to be considered as a two-body problem.  This perception continued into the era of quantum chromodynamics (QCD), with the pion being viewed as the simplest accessible manifestation of QCD dynamics and therefore the natural testing ground for theoretical methods that aim to elucidate a wide range of QCD phenomena.  Growing in parallel was an appreciation that the pion occupies a special place in nuclear and particle physics; viz., as the archetype for meson-exchange forces, and hence plays a critical role as an elementary field in the nuclear structure Hamiltonian \cite{Pieper:2001mp,Machleidt:2011zz}.  These conflicting views are reconciled in the modern paradigm \cite{Maris:1997hd}, which simultaneously describes the pion as a conventional bound-state in quantum field theory and the Goldstone mode associated with dynamical chiral symmetry breaking (DCSB).  This dichotomy entails that fine tuning cannot play any role in a veracious explanation of pion properties and ensures that elucidating the nature of its parton content is critical to any understanding of QCD.

One of the earliest predictions of the QCD parton model was the behaviour of the pion's valence-quark distribution function at large Bjorken-$x$ \cite{Ezawa:1974wm,Farrar:1975yb}:
$q^\pi(x) \sim (1-x)^2$.
Owing to the validity of factorisation in QCD, $q^\pi(x)$ is directly measurable in $\pi N$ Drell-Yan experiments.   
However, as described elsewhere \cite{Holt:2010vj}, conclusions drawn from a leading-order analysis of these experiments proved controversial, producing \cite{Conway:1989fs} $q^\pi(x) \sim (1-x)$ and thus an apparent disagreement with QCD.  We address this issue herein by first correcting a commonly used expression for the valence-quark distribution function and then illustrating its consequences with an algebraic model that incorporates salient features of QCD.

\medskip

\noindent\textbf{2.$\;$Quark distribution function in the pion}.
The hadronic tensor relevant to inclusive deep inelastic lepton-pion
scattering may be expressed in terms of two invariant structure functions \cite{Jaffe:1985je}.  In the deep-inelastic Bjorken limit \cite{Bjorken:1968dy}: $q^2\to\infty$, $P\cdot q \to -\infty$ but $x:= - q^2/[2 P \cdot q]$ fixed,
that tensor can be written $(t_{\mu\nu} = \delta_{\mu\nu}-q_\mu q_\nu/q^2, P_\mu^{\,t}= t_{\mu\nu}P_\nu)$
\begin{equation}
W_{\mu\nu}(q;P) = F_1(x)\, t_{\mu\nu} - \frac{F_2(x)}{P\cdot q}
\,P_\mu^{\,t} P_\nu^{\,t}\,,\;F_2(x) = 2 x F_1(x)\,.
\end{equation}
$F_1(x)$ is the pion structure function, which provides access to the pion's quark distribution functions:
\begin{equation}
\label{qPDF}
F_1(x) = \sum_{q\in \pi} \, e_q^2 \, q^\pi(x)\,,
\end{equation}
where $e_q$ is the quark's electric charge.  The sum in Eq.\,\eqref{qPDF} runs over all quark flavours; but in the $\pi^+$ it is naturally dominated by $u(x)$, $\bar d(x)$.  Moreover, in the $\mathpzc{G}$-parity symmetric limit, which we employ throughout, $u(x)=\bar d(x)$.  [Importantly, Bjorken-$x$ is equivalent to the light-front momentum fraction of the struck parton.]  The structure function may be computed from the imaginary part of the virtual-photon--pion forward Compton scattering amplitude: $\gamma(q) + \pi(P) \to \gamma(q) + \pi(P)$.

\begin{figure}[t]

\centerline{\includegraphics[width=0.5\linewidth]{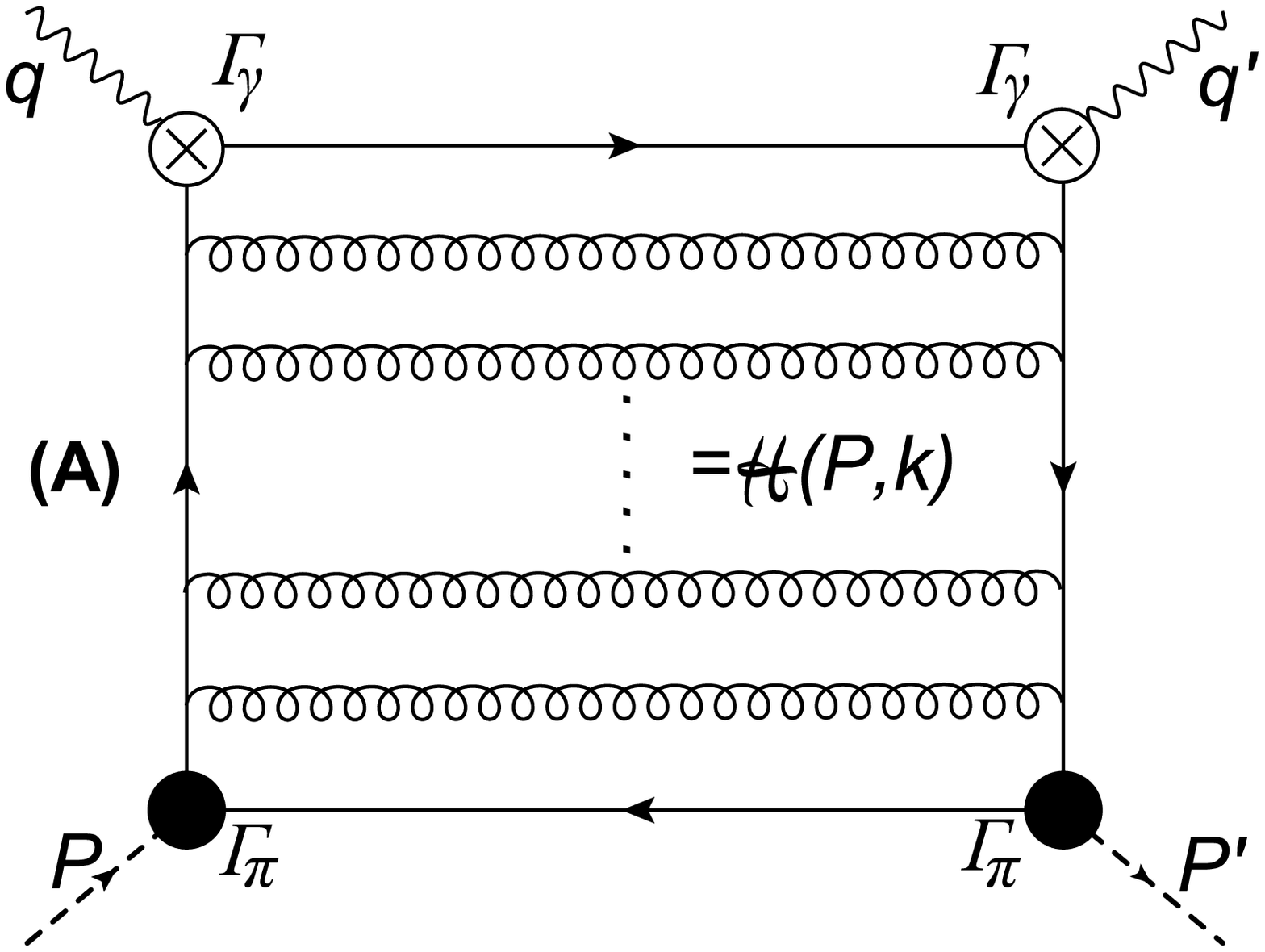}}

\medskip

\leftline{\includegraphics[width=0.45\linewidth]{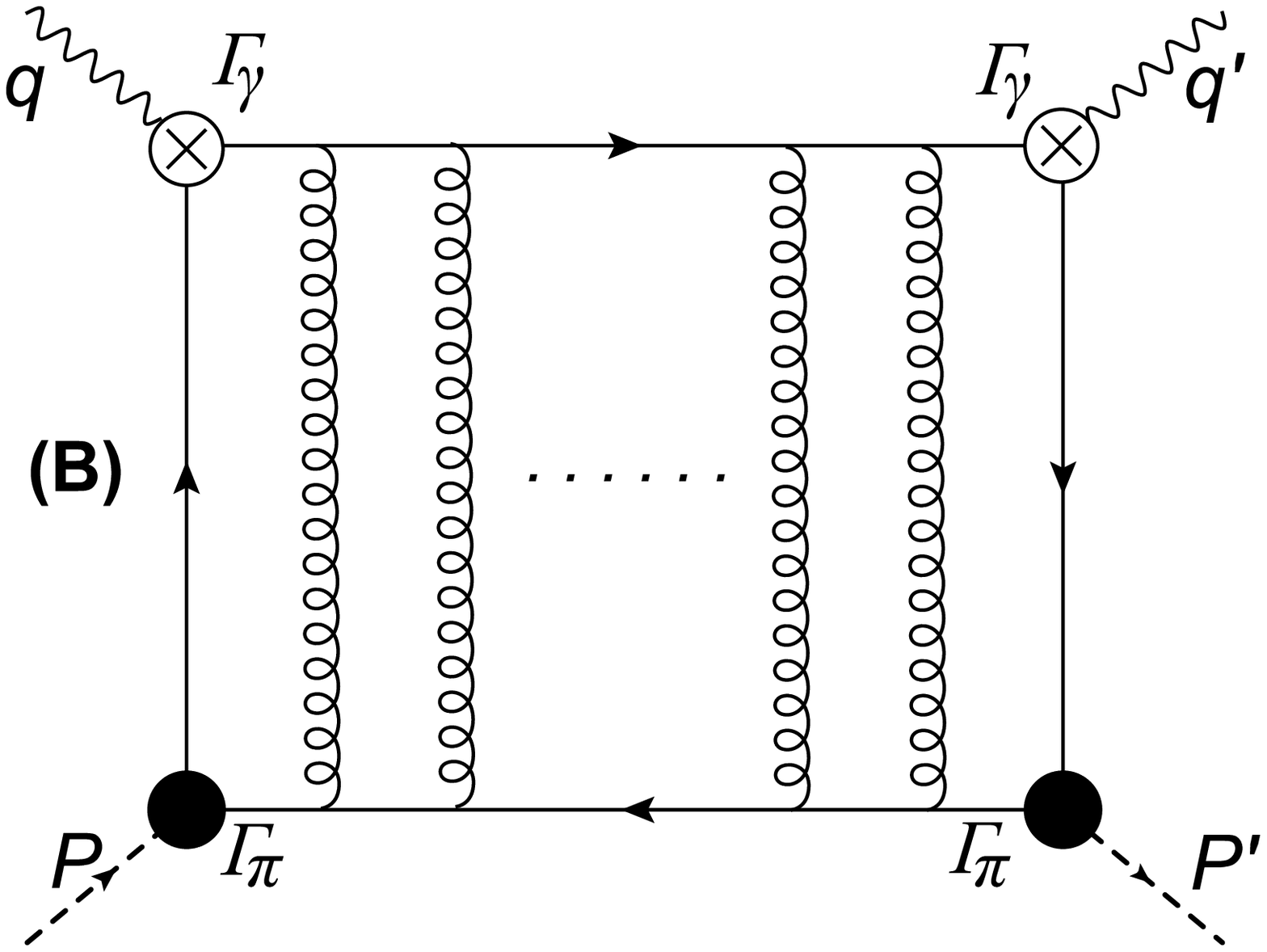}}

\vspace*{-19ex}

\rightline{\includegraphics[width=0.45\linewidth]{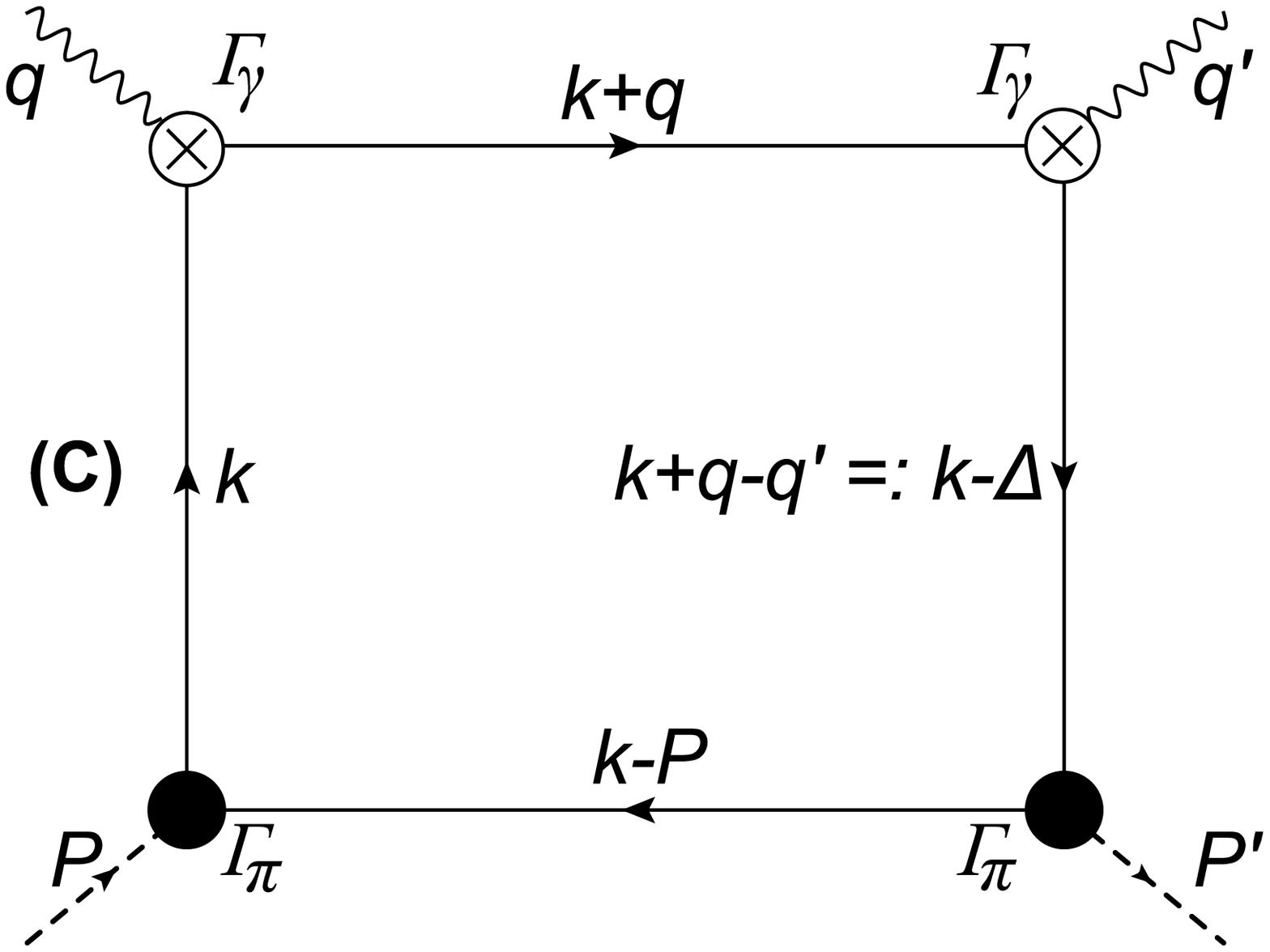}}

\caption{Amplitude-(1) for virtual-photon--pion Compton scattering in RL truncation is obtained from the sum $(A)+(B)-(C)$.  The ``dots'' in (A) and (B) indicate summation of infinitely many ladder-like rungs.
The other two amplitudes are obtained as follows: (2) -- switch vertices to which $q$ and $q^\prime$ are attached; and (3) -- switch vertex insertions associated with $q^\prime$ and $P^\prime$.
The lines and vertices mean the following:
\emph{dashed line} -- pion;
\emph{undulating line} -- photon;
\emph{spring} -- interaction-gluon in the RL kernels;
\emph{solid line} -- dressed-quark;
\emph{open-cross circle} -- dressed--quark-photon vertex;
\emph{filled circle} -- pion Bethe-Salpeter amplitude.  Each of the last three is computed in RL truncation.
\label{figCompton}}
\end{figure}

\medskip

\noindent\textbf{3.$\;$Rainbow-ladder truncation}.
Herein we analyse $q^\pi(x)$ in Eq.\,\eqref{qPDF} within the context of the rainbow-ladder (RL) truncation of QCD's Dyson-Schwinger equations \cite{Cloet:2013jya}.  That truncation is the leading-order term in a symmetry-preserving scheme \cite{Munczek:1994zz,Bender:1996bb,Chang:2009zb} which is accurate for, amongst other systems, isospin-nonzero-pseudoscalar-mesons because corrections in this channel largely cancel owing to parameter-free preservation of the Ward-Green-Takahashi (WGT) identities.

Following Ref.\,\cite{Bicudo:2001jqS}, it is evident that the virtual Compton amplitude in RL truncation should be built from permutations of the three diagrams illustrated in Fig.\,\ref{figCompton} \cite{Binosi:2003yf}.  This collection is necessary and sufficient to ensure preservation of the relevant WGT identities so long as the dressed-quark propagators, pion Bethe-Salpeter amplitudes and dressed--quark-photon vertices, appearing in the diagrams, are all computed in RL truncation.

Consider the virtual forward Compton amplitude in the Bjorken limit.  The Amplitude-(3) permutation of the diagrams in Fig.\,\ref{figCompton} corresponds to a collection of so-called \emph{cat's ears} contributions.  They are greatly suppressed compared to the other two permutations in the Bjorken limit and hence may be neglected.  The Amplitude-(2) permutation corresponds simply to symmetrising the incoming and outgoing photons and so need not explicitly be considered further.  Consequently, one may focus solely on those diagrams drawn explicitly in Fig.\,\ref{figCompton}.

In the forward and Bjorken limits, Diagram-(A) in Fig.\,\ref{figCompton} is the textbook \emph{handbag} contribution to virtual Compton scattering, which yields the following piece of the structure function:
\begin{equation}
\label{qvHklx}
q_{A}^\pi(x) = N_c{\rm tr} \! \int_{dk} \,
\delta_n^{x}(k_\eta)\, n\cdot \gamma \,{\cal H}_\pi(P,k)\,,
\end{equation}
where $N_c=3$ and the trace is over spinor indices; $\int_{dk} := \int \frac{d^4 k}{(2\pi)^4}$ is a translationally invariant regularisation of the integral; $\delta_n^{x}(k_\eta):= \delta(n\cdot k_\eta - x n\cdot P)$; $n$ is a light-like four-vector, $n^2=0$; $P$ is the pion's four-momentum, $P^2=-m_\pi^2$ and $n\cdot P = -m_\pi$, with $m_\pi$ being the pion's mass; and $k_\eta = k + \eta P$, $k_{\bar\eta} = k - (1-\eta) P$, $\eta\in [0,1]$.  Owing to Poincar\'e covariance, no observable can legitimately depend on $\eta$; i.e., the definition of the relative momentum.
Diagram-(A) is typically the only contribution retained in computations of the pion's quark distribution function; e.g., Refs.\,\cite{Shigetani:1993dx,Davidson:1994uv,Bentz:1999gx,%
Dorokhov:2000gu,Hecht:2000xa,Nguyen:2011jy}.

In RL truncation, ${\cal H}_\pi(P,k)$ is an infinite sum of ladder-like rungs, as illustrated in Fig.\,\ref{figCompton}, so that one may write \cite{Nguyen:2011jy}
\begin{align}
\nonumber
q_{A}^\pi(x)  = N_c  & {\rm tr} \! \int_{dk} \!
 i\Gamma_\pi(k_\eta,-P)\\
& \times \,S(k_\eta)\, \Gamma^n(k;x) \, S(k_\eta)\, i\Gamma_\pi(k_{\bar\eta},P)\, S(k_{\bar\eta})\,,
\label{Eucl_pdf_LR_Ward}\\
\label{DressedSk}
\rule{-0.8em}{0ex}\mbox{wherein} \rule{3em}{0ex} S(k) & = Z(k^2)/[i\gamma\cdot k + M(k^2)]
\end{align}
is the dressed-quark propagator, $\Gamma_\pi(k,P)$ is the pion's Bethe-Salpeter amplitude, and $\Gamma^n(k;x)$ is a generalisation of the quark-photon vertex, describing a dressed-quark scattering from a zero momentum photon.  It satisfies a RL Bethe-Salpeter equation with inhomogeneity $i n\cdot\gamma \,\delta_n^{x}(k_\eta)$ \cite{Nguyen:2011jy}.

This treatment of Diagram-(A) is precisely analogous to the symmetry preserving analysis of the pion's electromagnetic form factor (at $Q^2=0$) \cite{Maris:2000sk}.  Equation~\eqref{Eucl_pdf_LR_Ward} ensures \mbox{$ \int_0^1 dx \,  q_A^\pi(x) = 1$} because $ \int dx \, \Gamma^n(\ell;x)$ gives the Ward-identity vertex and the Bethe-Salpeter amplitude is canonically normalised.  The minimal \emph{Ansatz} sufficient to preserve these qualities is $\Gamma^n(k;x) =  \delta_n^{x}(k_\eta)\, \partial_{k_\eta} S^{-1}(k_\eta)$,
in which case one has
%
\begin{equation}
q_{A}^\pi(x) = N_c {\rm tr} \! \int_{dk}
\delta_n^{x}(k_\eta)\Gamma_\pi(k_\eta,-P) \partial_{k_\eta}S(k_\eta)\Gamma_\pi(k_{\bar\eta},P)\, S(k_{\bar\eta})\,.
\label{qAPDF}
\end{equation}

\begin{figure}[t]

\leftline{\includegraphics[width=0.45\linewidth]{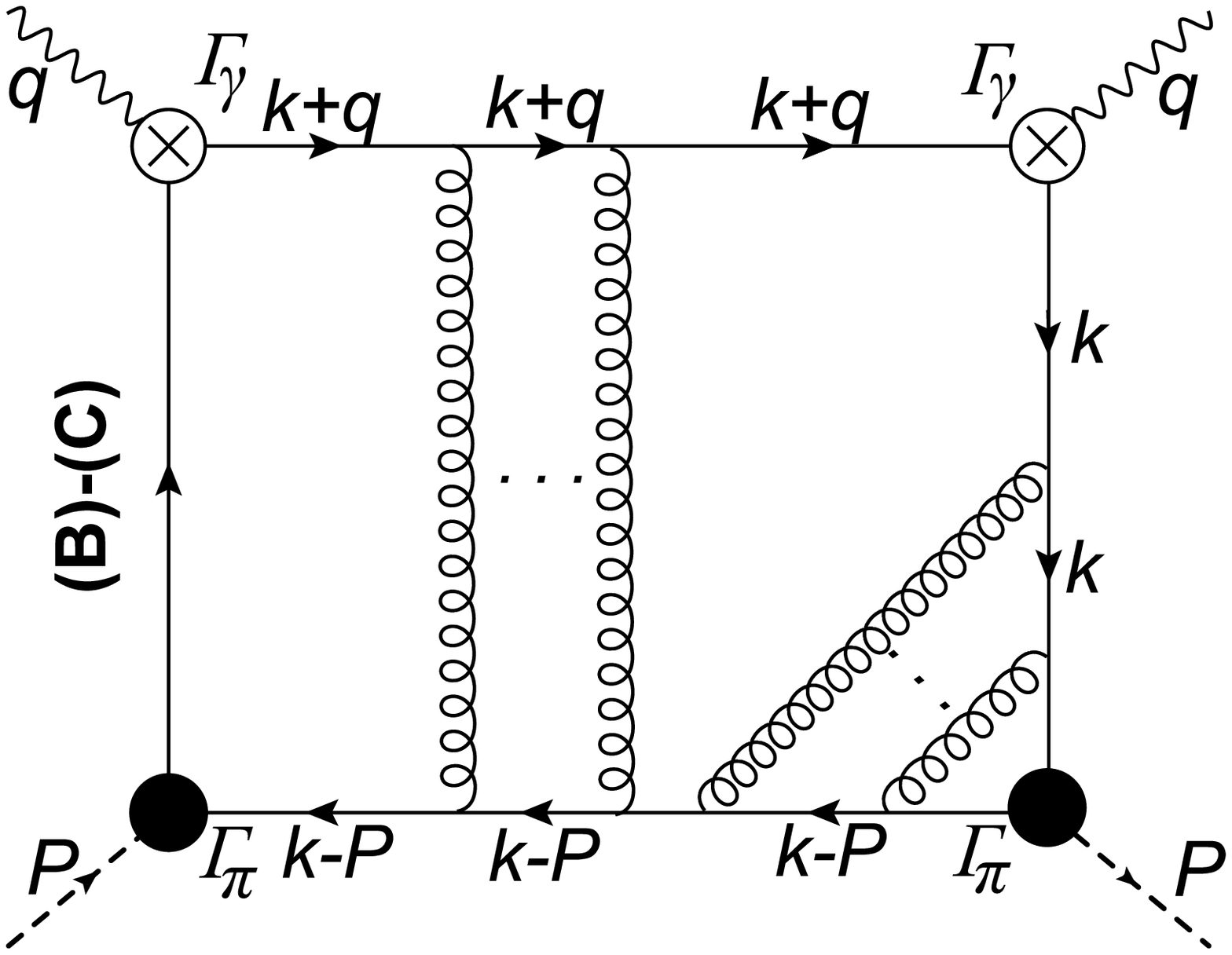}}
\vspace*{-17.3ex}

\rightline{\includegraphics[width=0.45\linewidth]{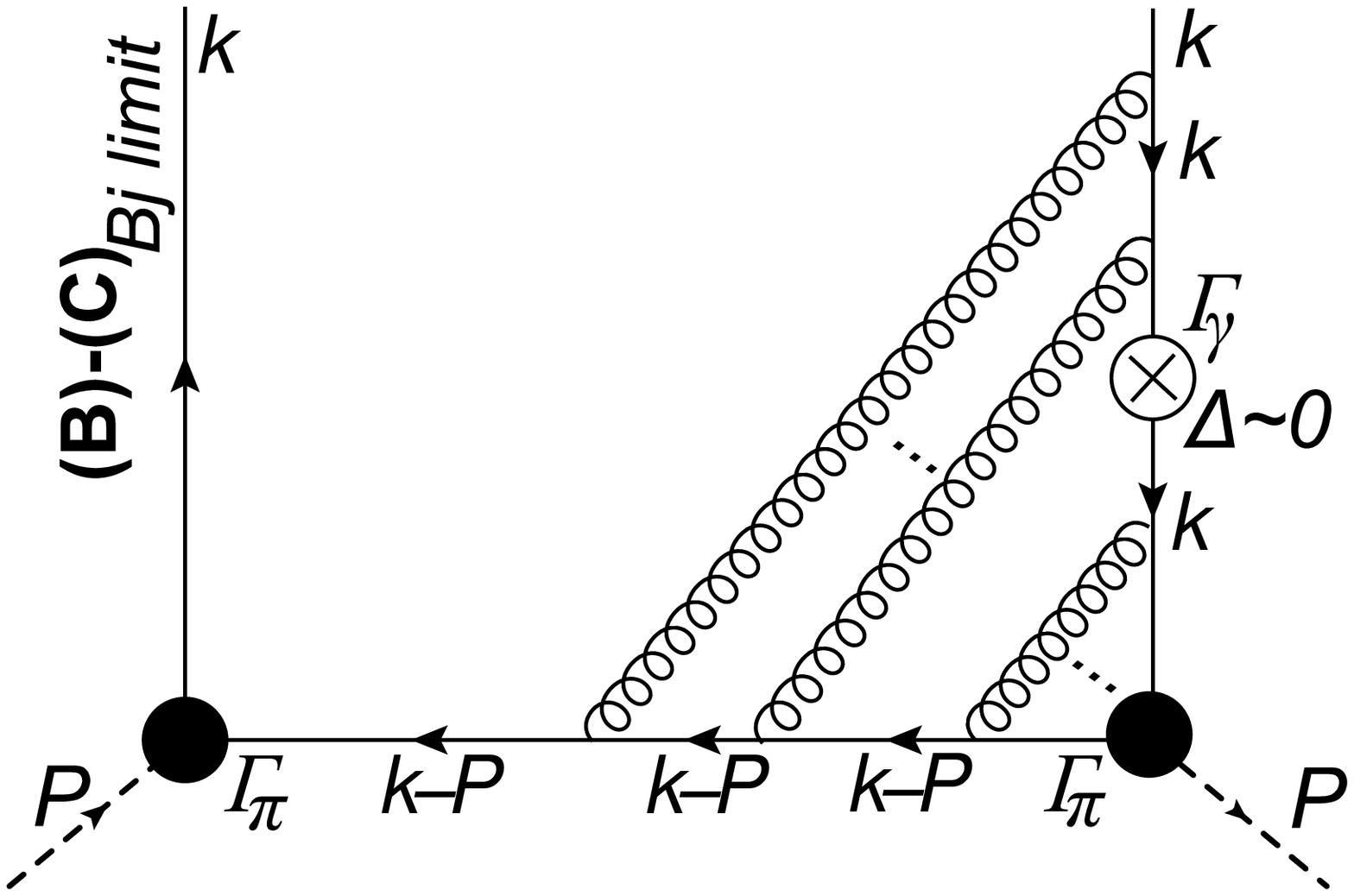}}

\caption{
\emph{Left panel} -- Forward limit of the combination $(B)-(C)$ in Fig.\,\ref{figCompton}.  The figure also exposes the internal structure of the pion's Bethe-Salpeter amplitude obtained in RL-truncation.
In the Bjorken limit, the initial/final-state interactions involve very soft gluons and hence, in combination with the ladder resummation, produce a contribution that is of the same order as Diagram-(A) in Fig.\,\ref{figCompton}.
\emph{Right panel} -- Imaginary part of the left panel in the Bjorken limit: the vertex insertion can appear between any two interaction lines.  The compound vertex on the right is readily simplified using the RL Bethe-Salpeter equation.
\label{figComptonBCF}}
\end{figure}

The other two diagrams in Fig.\,\ref{figCompton} have hitherto been overlooked. Given that the combination $(B)-(C)$ is crucial if the WGT identities are to be satisfied in a RL analysis of Compton scattering, it would seem a mistake to ignore these terms.  Let us therefore consider their content.  A first observation is that $(B)_0-(C)=0$; i.e., if one omits all terms from the ladder-like sum in Diagram-(B), then it is completely cancelled by Diagram-(C).  So, $(B)-(C)$ is a sum of infinitely many ladder-like rungs, beginning with one rung.  This is illustrated in Fig.\,\ref{figComptonBCF} (left panel), which also exposes the internal structure of the pion's RL-truncation Bethe-Salpeter amplitude.  Studying this figure, the nature of the combination $(B)-(C)$ becomes plain; viz., it expresses a photon being absorbed by a dressed-quark, which then proceeds to become part of the pion bound-state before re-emitting the photon.  Thinking perturbatively, one might imagine these processes to represent effects associated with initial/final-state interaction corrections to the handbag diagram and thus to be suppressed.  However, so long as the gluon exchanges are soft, which is the limit depicted in the left-panel of Fig.\,\ref{figComptonBCF}, that is not the case because the resummation of ladder-like rungs is resonant.  This contribution is thus of precisely the same order as that from Diagram-(A) in Fig.\,\ref{figCompton}.
In fact, akin to the final state interactions that produce single spin asymmetries \cite{Brodsky:2002cx}, the contribution we have identified is leading-twist and its appearance signals failure of the \emph{impulse approximation}.




To elucidate further, consider the imaginary part of Fig.\,\ref{figComptonBCF}--left-panel in the Bjorken limit, which produces the leading contribution illustrated in the right panel: the vertex insertion can appear between any pair of interaction lines.  Using the recursive structure of the ladder Bethe-Salpeter kernel and the Ward identity, which entails that inserting a zero-momentum vector-probe into a propagator line is equivalent to differentiation of the propagator, then the compound vertex on the right side of Fig.\,\ref{figComptonBCF}--right-panel is readily seen to correspond to differentiation of the Bethe-Salpeter amplitude itself with respect to $k_\eta$.  One thus arrives at the following contribution from $(B)-(C)$ to the pion's quark distribution function:
%
\begin{equation}
q_{BC}^\pi(x) = N_c {\rm tr}\! \int_{dk}\!
\delta_n^x(k_\eta)\partial_{k_\eta} \Gamma_\pi(k_\eta,-P)  S(k_\eta)\Gamma_\pi(k_{\bar\eta},P)\, S(k_{\bar\eta})\,.
\label{qBCPDF}
\end{equation}
This expression is nonzero in general.  It only vanishes when the pion's Bethe-Salpeter amplitude is independent of relative momentum; i.e., in the class of theories that employ a momentum-independent interaction, which includes models of the Nambu--Jona-Lasinio type \cite{Nambu:1961tp} and DSE-formulated analogues \cite{Roberts:2010rn}.

Adding Eqs.\,\eqref{qAPDF} and \eqref{qBCPDF}, we have our amended result for the quark distribution function in RL truncation:
\begin{subequations}
\label{TqFULL}
\begin{align}
\lefteqn{q_{L}^\pi(x) = q_{A}^\pi(x) + q_{BC}^\pi(x)}\\
%
%
 &= N_c  {\rm tr} \! \int_{dk}\!
\delta_n^x(k_\eta)\,
\partial_{k_\eta} \left[ \Gamma_\pi(k_\eta,-P)  S(k_\eta)\right] \Gamma_\pi(k_{\bar\eta},P)\, S(k_{\bar\eta})\,,
\label{qFULL}
\end{align}
\end{subequations}
where the derivative acts only on the bracketed terms.  Equation~\eqref{qFULL} is the minimal expression that retains the contribution to the quark distribution function from the gluons which bind dressed-quarks into the pion.  It produces results that are independent of $\eta$; i.e., the definition of the relative momentum.

\medskip

\noindent\textbf{4.$\;$Sketching the dressed-quark PDF}.
A range of novel insights into the dressed-quark structure of the pion can be obtained by using \cite{Chang:2013pqS}, with $\Delta_M(s)=1/[s+M^2]$,
\begin{subequations}
\label{NakanishiASY}
\begin{eqnarray}
\label{eq:sim1}
S(k) & = &[-i\gamma\cdot k+M]\Delta_M(k^2)\,,\\
\rho_\nu(z) & = & \frac{1}{\sqrt{\pi}}\frac{\Gamma(v+3/2)}{\Gamma(\nu+1)}(1-z^2)^\nu\,,\\
\label{eq:sim2}
\mathpzc{n}_\pi \Gamma_\pi(k;P) & = & i\gamma_5\int^1_{-1}dz\, \rho_\nu(z) \, \hat\Delta^\nu_M(k^2_{+z})\,,
\end{eqnarray}
\end{subequations}
where $M$ is a dressed-quark mass-scale; $\hat\Delta_M(s)=M^2\Delta_M(s)$; $k_{+z}=k+ z P/2$ and we work in the chiral limit $(P^2=0)$; and $\mathpzc{n}_\pi$ is the Bethe-Salpeter amplitude's normalisation constant.

To frame the analysis, one may begin by considering the pion's valence-quark parton distribution amplitude (PDA):
\begin{equation}
f_\pi\, \varphi_\pi(x) = N_c{\rm tr} \! \int_{dk} \!\!
\delta_n^x(k_\eta) \,\gamma_5\gamma\cdot n\, \chi_\pi(k;P)\,,
\label{pionPDA}
\end{equation}
where $\chi_\pi(k;P) = S(k_\eta) \Gamma_\pi(k;P) S(q_{\bar\eta})$ is the pion's Bethe-Salpeter wave function and $f_\pi$ is its leptonic decay constant.  A QCD-like theory corresponds to $\nu=1$ in Eq.\,\eqref{eq:sim2}, in which case Eq.\,\eqref{pionPDA} yields \cite{Chang:2013pqS}: $f_\pi\mathpzc{n}_\pi=N_c M/(8\pi^2)$; and
\begin{equation}
\label{fpival}
\varphi_\pi(x) = 6 x (1-x)=:\varphi^{\rm asy}(x)\,,
\end{equation}
i.e., the PDA appropriate to QCD's conformal limit \cite{Lepage:1979zb,Efremov:1979qk,Lepage:1980fj}.




Now consider $q_A^\pi(x)$ in Eq.\,\eqref{qAPDF}, which was hitherto the only contribution retained in evaluating the pion's dressed-quark distribution function.  There are numerous ways to evaluate the integrals that arise after inserting Eqs.\,\eqref{NakanishiASY}.  The simplest, perhaps, is to work with $\eta=0$, and use light-front coordinates and the residue theorem, thereby obtaining
%
%
\begin{align}
\nonumber & q_A^\pi(x) =
\mathpzc{n}_q
\left[x^3 (x [-2 (x-4) x-15]+30) \ln (x)+ (2 x^2+3)\right.\\
&\left. \times (x-1)^4 \ln (1-x)+x [x (x [2 x-5]-15)-3] (x-1)\right]\,,
\label{qAresult}
\end{align}
where $\mathpzc{n}_q = 9/(20\pi^2 \mathpzc{n}_\pi^2)$.  The result is independent of $\eta$, as one may establish by direct computation, and the $x$-dependence is independent of $M$, the defining mass-scale in Eqs.\,\eqref{NakanishiASY}. 
[Eq.\,\eqref{qAresult} has also been obtained via analysis of the pion's generalised parton distribution beginning with Eqs.\,\eqref{NakanishiASY} \cite{Mezrag}.]

Computation of $q_{BC}^\pi(x)$ in Eq.\,\eqref{qBCPDF} can similarly be completed:
%
%
%
\begin{align}
\nonumber
& q_{BC}^\pi(x) = \mathpzc{n}_q
\left[ x^3 (2 x ([x-3] x+5)-15) \ln (x) -(2 x^3+4 x+9) \right.\\
& \left. \times (x-1)^3 \ln (1-x) -x (2 x-1) ([x-1] x-9) (x-1)\right]\,.
\end{align}
The result is plainly nonzero; and it is also independent of $\eta$ and $M$. Given that this term was previously omitted, one must enquire into its importance.  The first thing to observe is
\begin{equation}
\mbox{$\int_0^1$} dx\, q_{BC}^\pi(x) = 0\,,
\end{equation}
so $q_{BC}^\pi(x)$ doesn't contribute net baryon number to the PDF.  One might have anticipated this from Fig.\,\ref{figComptonBCF} (right panel), which describes $q_{BC}^\pi(x)$ as adding momentum from the binding gluons.

In connection with baryon number then, only $q_{A}^\pi(x)$ can contribute; and,  as noted above, in a symmetry preserving analysis the normalisation of the Bethe-Salpeter amplitude ensures that the pion charge form factor is unity at $Q^2=0$ \cite{Roberts:1994hh}.  This condition is algebraically equivalent to $\int_0^1  dx\, q_{A}^\pi(x) = 1$,
so that Eqs.\,\eqref{NakanishiASY} are completed with $\mathpzc{n}_\pi^2 = 5/(32\pi^2)$. [A notion of scale is provided by the observation that $r_\pi^2 = 162/(125 M^2)$ and hence $M=0.33\,$GeV yields the empirical value  \cite{Beringer:1900zz} $r_\pi=0.67\,$fm.]

One is now in a position to consider the momentum sum rule; namely, to compute the light-front momentum fraction carried by the pion's dressed-quark in RL truncation:
\begin{equation}
\label{momsum}
\langle x \rangle_q^\pi = \int_0^1 dx\, \left[ x \, q_{A}^\pi(x)+ x\, q_{BC}^\pi(x)\right]
= \frac{117}{250} + \frac{8}{250} = \frac{1}{2}\,;
\end{equation}
viz., the dressed-quark and -antiquark each carry half the pion's momentum but that is only true after the leading contributions from all diagrams in Fig.\,\ref{figCompton} are summed.

\begin{figure}[t]

\centerline{\includegraphics[width=0.75\linewidth]{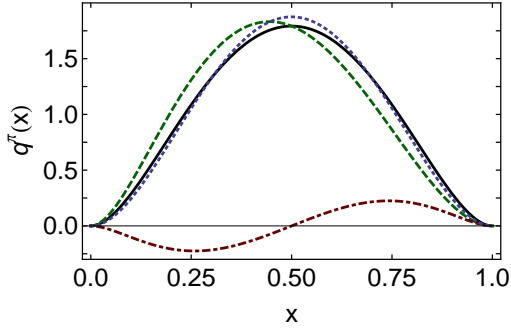}}

\caption{Pion dressed-quark distribution function in rainbow-ladder truncation: \emph{solid} -- complete result, Eq.\,\eqref{qRLSymmetric}; \emph{dashed} -- handbag contribution usually retained, Eq.\,\eqref{qAPDF}; and \emph{dot-dashed} -- amendment described in connection with Eq.\,\eqref{qBCPDF}.  An interesting comparison is provided by the \emph{dotted curve}: $q_{2}(x)=30 x^2(1-x)^2$.  To the eye, it is barely distinguishable from our complete result; and the mean value of the absolute relative difference between the curves is just 9\%.  Hence $q_2(x)$ can be useful as an approximation to Eq.\,\eqref{qRLSymmetric}.
\label{PDFplot}}
\end{figure}

Another important feature is hidden in Eq.\,\eqref{momsum}; namely, as illustrated in Fig.\,\ref{PDFplot}, including $q_{BC}^\pi(x)$ produces a symmetric dressed-quark PDF in RL truncation:
%
%
\begin{align}
\nonumber
q_{L}^\pi(x) & =
    \frac{72}{25} \left[ x^3 (x [2 x-5]+15) \ln (x) + (x [2 x+1]+12) \right.\\
& \left. \times (1-x)^3 \ln (1-x) +2 x (6-[1-x] x) (1-x)\right]\,.
\label{qRLSymmetric}
\end{align}
This is logical because the dressed-quark and -antiquark are the sole measurable constituents of the pion in an internally consistent RL computation: they and their associated bound-state amplitude absorb and contain all contributions from sea or glue partons.  It follows that if the dressed quark carries a fraction $x$ of the pion's momentum, the dressed antiquark carries $[1-x]$.  Notably, only a symmetric distribution produces $\langle x \rangle_q^\pi=1/2$ without fine-tuning.

We close this part with an analysis of the large-$x$ behaviour.  As noted in connection with Eq.\,\eqref{fpival}, QCD-like scaling behaviour is obtained with $\nu=1$.  It is thus unsurprising that
\begin{equation}
\label{valencepower}
q_{L}^\pi(x) \stackrel{x\sim 1}{=} \frac{216}{5}\, (1-x)^2 + \mbox{O}([1-x]^3)\,.
\end{equation}
This is the power-law predicted by the QCD parton model \cite{Ezawa:1974wm,Farrar:1975yb}, obtained simply and exactly.  Owing to symmetry under $x\leftrightarrow [1-x]$, the same power-law is manifest on $x\sim 0$, a result which emphasises that $q_{L}^\pi(x)$ is truly a constituent-like distribution: any sea-quark contamination would produce a marked asymmetry.  Notably, there is a direct connection between the $k^2$-dependence of a theory's interaction [$\nu$ in Eq.\,\eqref{eq:sim2}], the behaviour of the asymptotic PDA ($\varphi_\nu^{\rm asy}\propto [x(1-x)]^\nu$ \cite{Chang:2013pqS}) and the PDF's power-law behaviour at a characteristic hadronic scale, $\zeta_H$ [defined below]:
\begin{equation}
\forall\nu \geq 0\,: q_{L}^\pi(x) \approx [\varphi_\nu^{\rm asy}(x)]^2
\Rightarrow q^\pi(x;\zeta_H) \stackrel{x\simeq 1}{\sim} (1-x)^{2\nu}  \,.
\end{equation}

\medskip

\noindent\textbf{5.$\;$Incorporating sea-quarks and glue}.
The dressed-quark basis obtained using the rainbow-ladder truncation with a realistic one-loop renormalisation-group-improved (RGI) interaction \cite{Qin:2011dd} provides a good description of a wide range of pion properties \cite{Cloet:2013jya}, including its mass and decay constant, electroweak form factors, and $\pi\pi$ scattering.  It produces the dressed-quark PDF in Eq.\,\eqref{qFULL}, which is invariant under $x\leftrightarrow [1-x]$, and consequently generates a purely valence-quark distribution.  As is evident from the illustrations in Sect.\,3, this is because RL truncation includes no mechanism that can shift momentum from the dressed-quarks into sea-quarks and gluons: a RL pion is constituted solely from a dressed-quark and dressed-antiquark.  [This explains the useful feature that whilst Eqs.\,\eqref{NakanishiASY} produce the conformal-limit PDA from the exact formula in Eq.\,\eqref{pionPDA}, they cannot generate the asymptotic valence-quark PDF, $q^\pi(x) = \delta(x)$, from the approximation in Eq.\,\eqref{TqFULL}.]


In the context of the pion's PDFs, corrections to the RL truncation can be separated into two classes: [C1] redistributes baryon-number and momentum into the dressed-quark sea; and [C2] shifts momentum into the dressed-gluon distribution within the pion.  Perhaps the most obvious contributions within [C1] are those associated with what have been \emph{called resonant} or \emph{meson-cloud} corrections to the kernels in the gap and scattering equations.  One example is
\begin{equation}
\pi^+ = u \bar d \to u (\bar d d) \bar d = (u \bar d) (d \bar d) \sim \pi^+ \rho^0 \to u \bar d = \pi^+,
\end{equation}
which describes a RL-$\pi^+$, dressing itself with a RL-$\rho^0$.  This process enables the photon to interact with RL-dressed $\bar u$- and $d$-quark components within the physical $\pi^+$, thereby shifting momentum into the pion's RL-dressed sea.  Let us associate a total flux ``$Z$'' with such fluctuations.  In a symmetry preserving treatment, such processes do not change the total baryon-number content of the pion but they do reduce the probability of finding the RL-pion within the physical pion; and hence the quark distribution becomes
\begin{equation}
q_{vs}^\pi(x) = (1-Z) \, q_{L}^\pi(x) + Z \, q_{M}^\pi(x)\,,\;
\mbox{$\int_0^1dx\,q_{M}^\pi(x)=1$}\,,
\end{equation}
where $q_{M}^\pi(x)$ describes the cumulative effect on the PDF of all resonant corrections to the RL computation. 

\begin{figure}[t]

\leftline{\includegraphics[width=0.45\linewidth]{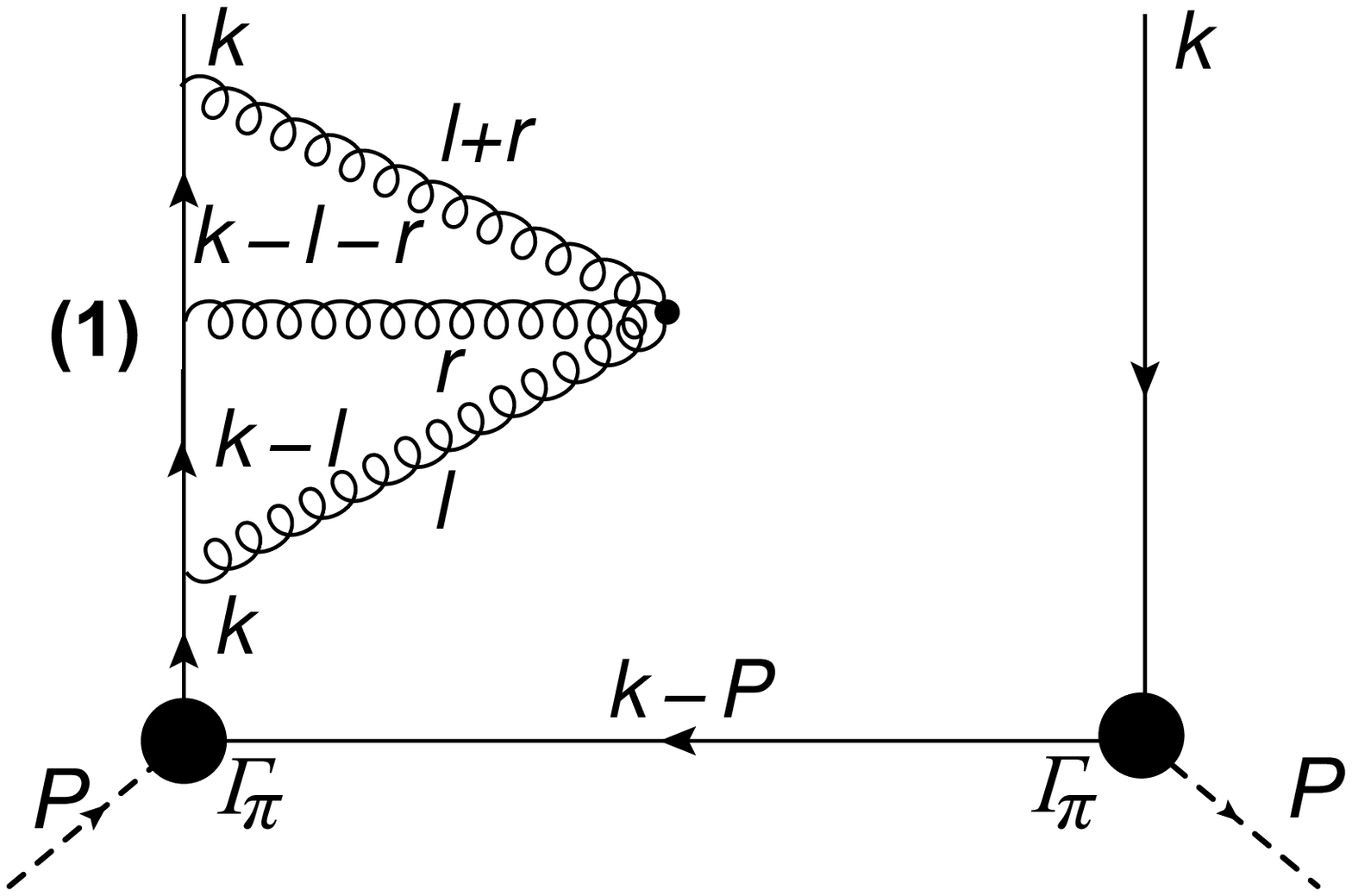}}
\vspace*{-17.3ex}

\rightline{\includegraphics[width=0.45\linewidth]{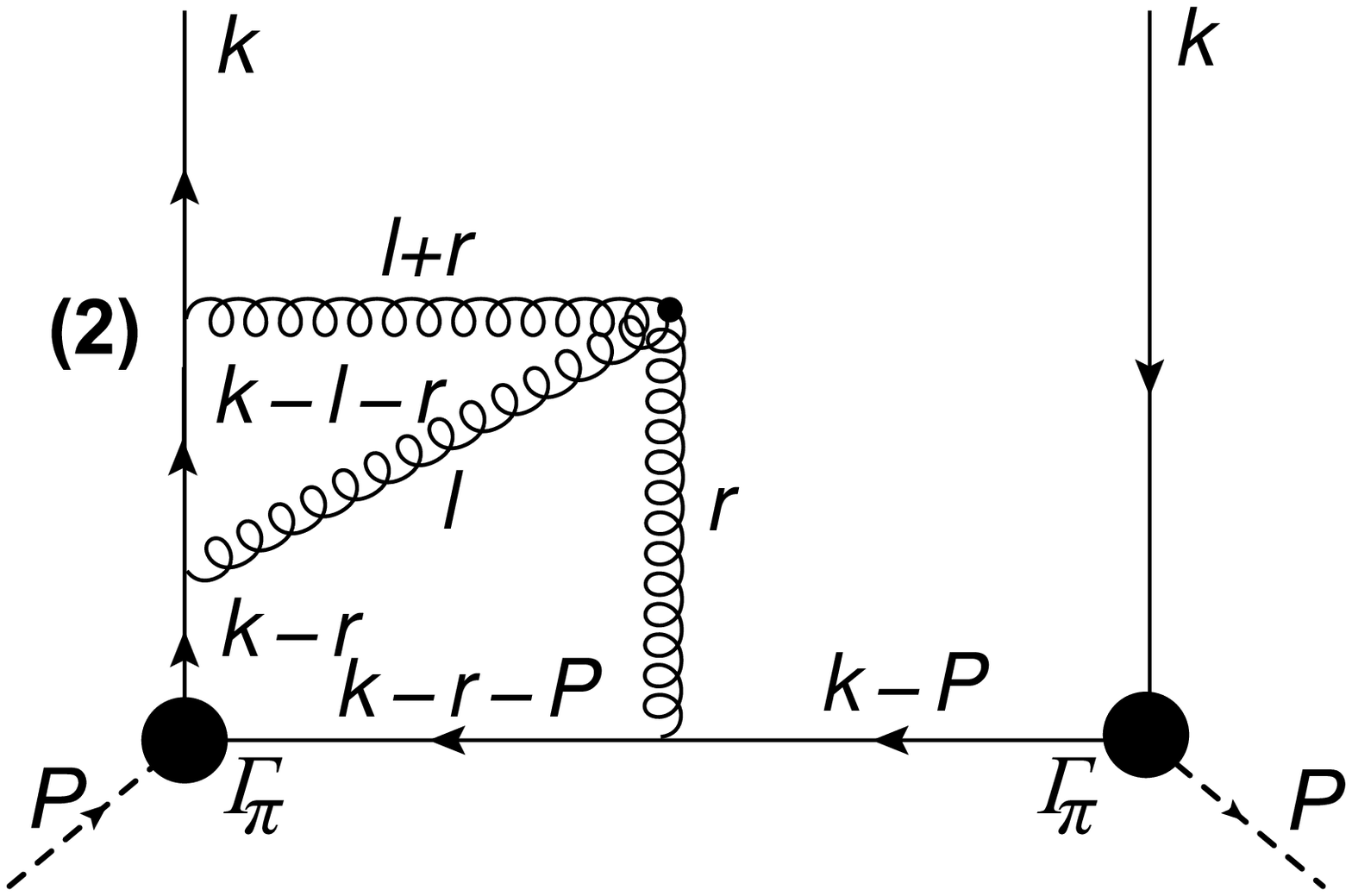}}

\centerline{\includegraphics[width=0.5\linewidth]{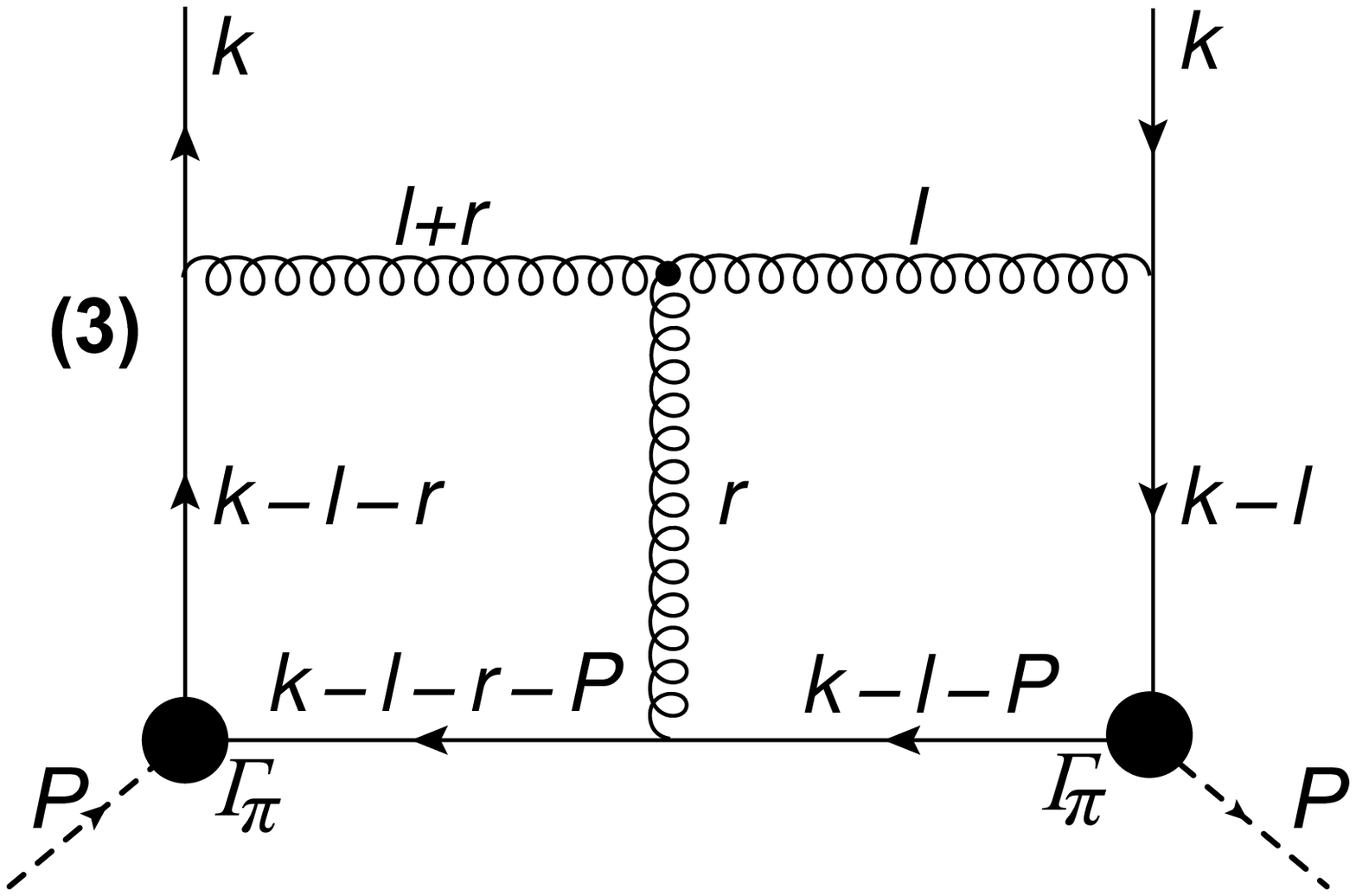}}

\caption{Illustration of some corrections to RL truncation.  Effects such as those illustrated in Diagrams-(1) and -(2) can be absorbed into the dressed-quark and the pion's Bethe-Salpeter amplitude and belong in [C1]; but Diagram-(3) is [C2] because it shifts momentum into the pion's dressed-gluon distribution without altering the distribution of baryon number
\label{figClass2}}
\end{figure}

In order to identify the second class of corrections, [C2], consider Fig.\,\ref{figClass2}.  Contributions of the type in Diagram-$(1)$ add additional dressing to the quark; and those in Diagram-$(2)$ add additional binding within the pion's Bethe-Salpeter amplitude.  As such, they can be absorbed into the distribution of the dressed-quarks and dressed-antiquarks within the pion, so that they serve mainly to modify the natural basis states and have little noticeable impact on either $q_{L}^\pi(x)$ or $q_{M}^\pi(x)$.  On the other hand, Diagram-$(3)$ has the effect of shifting momentum into the pion's gluon distribution.  Evidently, whilst the struck dressed-quark is still carrying a fraction $x$ of the pion's momentum, the momentum of the spectator system is shared between the dressed-antiquarks and -gluons: attributing a net $x_g>0$ to the dressed-gluon, then the dressed-antiquark carries $1-x-x_g$.  In a symmetry preserving treatment, corrections in [C2] have no impact on net baryon number within the pion but they do rob momentum from the baryon-number-carrying dressed-partons; namely, $q_{L,M}^\pi(x) \to q_{L_g,M_g}^\pi(x)$ with
\begin{subequations}
\label{DefineShift}
\begin{eqnarray}
\mbox{$\int_0^1$} dx \,q_{L_g,M_g}^\pi(x) \;\; &=& \mbox{$\int_0^1$} dx \,q_{L,M}^\pi(x)\,,\\
\mbox{$\int_0^1$} dx \, x\,q_{L_g,M_g}^\pi(x) &<& \mbox{$\int_0^1$} dx \, x\, q_{L,M}^\pi(x)\,.
\end{eqnarray}
\end{subequations}
Thus, with $\delta_g q_{L,M}^\pi(x) := q_{L_g,M_g}^\pi(x)-q_{L,M}^\pi(x)$, one arrives finally at the complete dressed-quark distribution function
\begin{equation}
\label{qPDFfinal}
q^\pi(x) = (1-Z)[q_{L}^\pi(x) + \delta_g q_{L}(x)]
+ Z[q_{M}^\pi(x) + \delta_g q_{M}(x)] \,.
\end{equation}

A procedure one may follow in order to compute the pion's valence-quark distribution function, Eq.\,\eqref{qPDFfinal}, is now apparent: begin with RL results, obtained using a sophisticated RGI kernel and with the resolution set via renormalisation at a particular scale $\zeta_{\rm H}$; then proceed systematically to add the corrections identified above; and, finally, use DGLAP evolution \cite{Dokshitzer:1977,Gribov:1972,Lipatov:1974qm,Altarelli:1977} to obtain the result at any other scale $\zeta>\zeta_{\rm H}$.  The last step is simply a labour-saving device because it eliminates the need for complete recomputation of the PDF at the new scale.  In this way one fixes \emph{a priori} that parameter, $\zeta_{\rm H}$, which practitioners usually identify as the \emph{typical hadronic scale}, 
and whose variation provides them with considerable flexibility as they seek to validate their model through a fit to data.

It is natural to ask for the value of $\zeta_{\rm H}$ at which the RL result alone should be most realistic.  That is $\zeta_{\rm H}\simeq 0\,$GeV, because the light-front momentum fraction carried by dressed-sea and -glue diminishes as $\zeta$ is reduced.  However, use of the available DGLAP equations at such a small value of $\zeta_{\rm H}$ is impossible because they are only valid on the perturbative domain.  What, then, is a suitable compromise?  An answer was provided in Ref.\,\cite{Holt:2010vj}: one should use $\zeta_{\rm H} \geq 2\Lambda_{\rm QCD} \approx 0.5\,$GeV, which corresponds to a scale whereat the chiral-limit dressed-quark mass-function, $M(p^2)$ in Eq.\,\eqref{DressedSk}, is concave-up (convex) and dropping rapidly but does not yet exhibit the behaviour associated with its truly asymptotic momentum-dependence.  As explained elsewhere \cite{Chang:2012rk}, it is only for momenta within this domain that a rigorous connection with perturbative QCD (pQCD) exists: it is impossible to begin at a smaller scale because then the crucial elements in any calculation, e.g., the dressed-quark propagator, exhibit momentum dependence that is essentially nonperturbative in origin, such as the inflexion point associated with confinement \cite{Cloet:2013jya}.  Notably, the expansion parameter in the DGLAP equations is $\alpha(s)/[2\pi]$, where $\alpha(s)$ is the strong running coupling; and  $\alpha(4\Lambda_{\rm QCD}^2)/(2\pi)\approx 0.17$ whereas $\alpha(2\Lambda_{\rm QCD}^2)/(2\pi)\approx 0.34$, which further vitiates any choice $\zeta_{\rm H}<2\Lambda_{\rm QCD}$.


Some remarks are in order before proceeding.  Notwithstanding the existence of calculable corrections to the RL truncation, the dressed-quarks and bound-states obtained in RL truncation provide a good basis for describing numerous hadron observables.  This is readily illustrated via the pion's electromagnetic form factor, $F_\pi(Q^2)$.  Meson-loop corrections only measurably affect its low-$Q^2$ behaviour, contributing $\lesssim 15\%$ to $r_\pi^2$ (squared-charge-radius) \cite{Alkofer:1993gu}; and gluonic corrections analogous to Diagram-(3) serve only to modify the form-factor's anomalous dimension \cite{Lepage:1980fj,Maris:1998hc,Cloet:2013jya}.  The salient features of $F_\pi(Q^2)$, including parton model scaling and the existence of scaling violations, are captured by the RL truncation \cite{Chang:2013nia}.

\medskip

\noindent\textbf{6.$\;$Illustrating the essentials}.
Equations~\eqref{TqFULL} and the framework in Sect.\,4 can be used to illustrate what may reasonably be expected from the procedure described in Sect.\,5.  This is valuable because, as will become apparent, differences between this illustration and results obtained using the complete procedure cannot be qualitatively significant.

The illustration can be built upon two observations, one concerning the dressed-sea distribution and the other relating to glue.  Note first that with realistic masses, meson-loop corrections to the RL result for the pion electromagnetic form factor at $Q^2=0$ are an O(5\%) effect.  This is evident in Ref.\,\cite{Alkofer:1993gu} and also in the result that, absent chiral corrections, the pion's leptonic decay constant is \cite{Gasser:1983yg} $f_0^2 \approx (0.09\,{\rm GeV})^2$ cf.\ experiment \cite{Beringer:1900zz}, $f_\pi^2 \approx (0.092\,{\rm GeV})^2$.  In Eq.\,\eqref{qPDFfinal}, one may therefore fix
\begin{equation}
\label{Zvalue}
Z=0.05\,.
\end{equation}
Regarding the profile of the dressed-sea contribution, we draw guidance from empirical information on $\pi N$ Drell-Yan \cite{Gluck:1999xe}:
\begin{equation}
\label{GRSsea}
x q_M^\pi(x) = \frac{1}{\mathpzc{N}} x^{\bar\alpha} (1-x)^{\bar\beta} (1 - \bar\gamma \sqrt{x} + \bar\delta_ x)
\end{equation}
where $1/\mathpzc{N}$ is a simple algebraic factor that ensures $\int_0^1dx\,q_M^\pi(x)=1$.
Then, at $\zeta_{\rm H}=0.51\,$GeV an empirical assessment of the pion's sea-quark distribution is consistent with
\begin{equation}
\label{SeaParameters}
\bar\alpha=0.16\,,\; \bar\beta=5.20\,,\;\bar\gamma=3.243\,,\;\bar\delta=5.206\,.
\end{equation}

The same consideration of $\pi N$ Drell-Yan shows that 29\% of the pion's momentum is carried by glue at $\zeta_{\rm H}$ $[\langle x_g\rangle =0.29]$, in a distribution that has \cite{Gluck:1999xe} $\alpha_g \approx 3/2$ and $\beta_g \approx 1+\beta_V$, where $\beta_V$ is the exponent which characterises the pion's valence-quark distribution on $x\simeq 1$.  In Eq.\,\eqref{qPDFfinal}, we therefore use $\delta_g q_{L,M} = \delta_g q$,
\begin{equation}
\label{shift}
\delta_g q(x) = s_g \, x^{\alpha_g-1} (1-x)^{\beta_g} P_1^{(\beta_g \, \alpha_g)}(2x-1)\,,
\end{equation}
with $s_g$ a parameter, in order to shift 29\% of the RL-dressed quarks' momentum into the gluon distribution.  [Equation~\eqref{shift} is consistent with Eqs.\,\eqref{DefineShift}.]  With $\beta_g=3$, owing to Eq.\,\eqref{valencepower}, one finds $s_g=8.5$.  [This procedure and Eq.\,\eqref{shift} are suggested by the dot-dashed curve in Fig.\,\eqref{PDFplot}, which shows how the resummation of gluon lines into the Bethe-Salpeter amplitude effects a redistribution of the dressed-quark momentum.]

Using Eqs.\,\eqref{qRLSymmetric} and \eqref{qPDFfinal}--\eqref{shift}, the pion's dressed-quark distribution function is completely determined.  Notably, it is not very sensitive to the values of the $\langle x_s\rangle \approx Z$ and $\langle x_g\rangle $, so long as $\langle x_s+x_g\rangle$, the momentum fraction contained in sea-quarks and glue, remains constant.  We have explained our preferred values of $Z$, $\langle x_g\rangle $.  Given the simplicity of our input, there is little sense in fine tuning them; but we will subsequently illustrate the effect of increasing the sum to $\langle x_s+x_g\rangle=0.40$ at $\zeta_H$.

Before proceeding further, however, it is worth comparing our model's predictions with available results from numerical simulations of lattice-regularised QCD (lQCD).  Such studies typically work with a resolving scale $\zeta_2 = 2\,$GeV, so comparison requires DGLAP evolution of our prediction from $\zeta_{\rm H} \to \zeta_2$.  That is readily accomplished by working with the Mellin moments.  One first computes
\begin{equation}
\label{LOevolution}
\langle x_{\zeta_{\rm H}}^m \rangle_q^\pi = \mbox{$\int_0^1$}dx \, x^m q^\pi(x;\zeta_{\rm H})
\end{equation}
up to a maximum number of moments: we use $m_{\rm max}=40$.  Then, at $\zeta>\zeta_{\rm H}$ \cite{Gluck:1995yr}:
\begin{equation}
\langle x_\zeta^m \rangle_q^\pi =\langle x_{\zeta_{\rm H}}^m \rangle_q^\pi
\left[\frac{\alpha(\zeta^2)}{\alpha(\zeta_{\rm H}^2)}\right]^{\gamma_0^m/\beta_0},
\end{equation}
where $\beta_0=11 - (2/3)n_f$,
\begin{equation}
\gamma_0^m = - \frac{4}{3} \left[ 3 + \frac{2}{(m+1)(m+2)}
- 4 \sum_{k=1}^{m+1} \frac{1}{k}\right]\,.
\end{equation}
[We use \cite{Qin:2011dd} $n_f=4$, $\Lambda_{\rm QCD}=0.234\,$GeV in the computation.]  An approximation to $q^\pi(x;\zeta_2)$ is readily reconstructed from the evolved moments by supposing the PDF is well described by a distribution with the functional form in Eq.\,\eqref{GRSsea}.  The procedure yields
\begin{equation}
q^\pi(x;\zeta_2) = 
4.21 x^{0.18} (1-x)^{2.10} .
\end{equation}
Equation~\eqref{LOevolution} describes leading-order evolution.  Any material differences generated by next-to-leading-order (NLO) evolution are masked by a 25\% increase in $\zeta_{\rm H}$ \cite{Gluck:1999xe}.

Owing to the loss of Poincar\'e-covariance, existing lQCD algorithms only provide access to the lowest three nontrivial moments of $q^\pi(x)$.  A contemporary simulation \cite{Brommel:2006zz}, using two dynamical fermion flavours, $m_\pi \gtrsim 0.34\,$GeV and nonperturbative renormalisation at $\zeta_2=2\,$GeV, produces the first row here:
\begin{equation}
\begin{array}{l|lll}
    & \langle x \rangle & \langle x^2 \rangle & \langle x^3 \rangle\\\hline
\mbox{\cite{Brommel:2006zz}} & 0.27(1) & 0.13(1) & 0.074(10)\\
\mbox{\cite{Best:1997qp}} & 0.28(8) & 0.11(3) & 0.048(20)\\
\mbox{\cite{Detmold:2003tm}} & 0.24(2) & 0.09(3) & 0.053(15)\\
{\rm average} & 0.26(8) & 0.11(4) & 0.058(27)\\\hline
{\rm herein} & 0.28 & 0.11 & 0.057
\end{array}\,.
\end{equation}
The results in Ref.\,\cite{Brommel:2006zz} agree with those obtained in earlier estimates based on simulations of quenched lQCD \cite{Best:1997qp,Detmold:2003tm} and are consistent with the values obtained from our computed distribution, evolved to $\zeta_2$.

\begin{figure}[t]

\centerline{\includegraphics[width=0.83\linewidth]{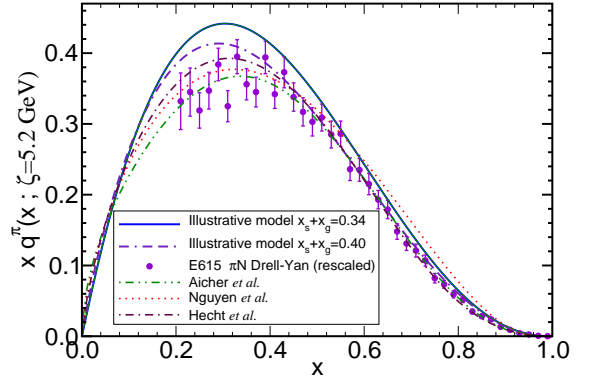}}

\caption{Pion dressed-quark distribution function.  \emph{Solid curve} -- result obtained herein with $\langle x_s+x_g\rangle =0.34$ and \emph{dot-dash-dash} -- result obtained with $\langle x_s+x_g\rangle=0.40$, illustrating the effect of shifting 10\% more of the dressed-quarks' momentum into sea and glue; data -- Ref.\,\cite{Conway:1989fs}, rescaled according to the reanalysis described in Ref.\,\cite{Aicher:2010cb} (\emph{dot-dot-dashed curve}); \emph{dotted} -- DSE result in Ref.\,\cite{Nguyen:2011jy}; and \emph{dot-dashed} -- first DSE prediction \cite{Hecht:2000xa}.
\label{figxqvx}}
\end{figure}

In Fig.\,\ref{figxqvx} we compare our result with available experiment \cite{Conway:1989fs}.  The average mass-scale for the data is $\zeta_5=5.2\,$GeV \cite{Wijesooriya:2005ir}, so we plot our prediction for $x q^\pi(x)$ evolved from $\zeta_{\rm H} \to \zeta_5$; viz.,
\begin{equation}
q^\pi(x;\zeta_5)=3.47 x^{0.021} (1-x)^{2.33} \,.
\end{equation}
In considering the data in Fig.\,\ref{figxqvx}, it is important to recall that E615 \cite{Conway:1989fs} reported a PDF obtained via a leading-order analysis in pQCD; and, as noted in Sect.\,1 and discussed elsewhere \cite{Hecht:2000xa,Wijesooriya:2005ir,Holt:2010vj}, this yielded controversial behaviour on $x\simeq 1$, contradicting QCD-based expectations: producing $q^\pi(x) \sim (1-x)$ instead of $q^\pi(x)\sim(1-x)^2$.  A subsequent NLO reanalysis \cite{Aicher:2010cb}, which, crucially, also included soft-gluon resummation, indicated that the data are actually consistent with $q^\pi(x)\sim(1-x)^2$: as emphasised by Ref.\,\cite{Wijesooriya:2005ir}, NLO evolution alone cannot expose that.  Thus, in Fig.\,\ref{figxqvx} we plot the E615 data rescaled as follows ${\rm E615}_{\rm 2010} = \mathpzc{F}(x)\,{\rm E615}_{\rm 1989}$, where $\mathpzc{F}(x)$ is the $x$-dependent ratio of Fit-3 in Ref.\,\cite{Aicher:2010cb} to the E615 fit described in Table~VII of Ref.\,\cite{Sutton:1991ay}.  It is apparent in Fig.\,\ref{figxqvx} that the data and all QCD-based calculations agree on the behaviour of $q^\pi(x)$ within the valence-quark domain.  

\medskip

\noindent\textbf{7.$\;$Conclusions and prospects}.
A useful starting point for the analysis of parton distribution functions and amplitudes is the rainbow-ladder (RL) truncation of QCD's Dyson-Schwinger equations.  This framework provides a description of hadrons via a dressed-quark basis, the accuracy of which in any given channel is knowable \emph{a priori}.  In this connection, we argued that the impulse-approximation expression used hitherto to define the pion's dressed-quark distribution function is incorrect owing to omission of contributions from the gluons which bind dressed-quarks into the pion.  The corrected expression [Eq.\,\eqref{TqFULL}] ensures that, independent of model details,  RL-dressed quarks define a purely valence distribution, they always each carry one-half of the pion's light-front momentum [Eq.\,\eqref{momsum}], and the valence-quark distribution behaves as $(1-x)^2$ on $x\simeq 1$ [Eq.\,\eqref{valencepower}].  Using algebraic formulae for the dressed-quark propagator and pion Bethe-Salpeter amplitude, which express effects associated with dynamical chiral symmetry breaking and produce the correct asymptotic pion parton distribution amplitude, we computed the RL result for the pion's valence-quark momentum distribution function [Fig.\,\ref{PDFplot}].

We subsequently explained [Sect.\,5] that corrections to the RL prediction for the pion's structure function may be divided into two classes: [C1], which redistributes baryon-number and momentum into the dressed-quark sea; and [C2], which shifts momentum into the pion's dressed-gluon distribution.  So far as one can determine empirically, contributions within [C2] are most important at an hadronic scale; viz., $\zeta_{\rm H} \approx 2\,\Lambda_{\rm QCD}$.  Working with this information, we built a simple algebraic model to express the principal impact of both classes of corrections, which, coupled with the RL prediction, permitted a realistic comparison with existing experiment [Fig.\,\ref{figxqvx}].  This enabled us to reveal essential features of the pion's valence-quark distribution.  Namely, at a characteristic and reasonable hadronic scale,
the pion's valence-quark distribution behaves as $(1-x)^2$ for $x\gtrsim 0.85$; %
and the valence-quarks carry roughly two-thirds of the pion's light-front momentum. 
It follows from this analysis that extant measurements of the pion's valence-quark distribution function confirm basic features of QCD.

On the other hand, a valuable opportunity is now available.  Employing the methods introduced in Refs.\,\cite{Chang:2013pqS,Chang:2013nia,Chang:2013epa}, one can follow the procedures sketched in Sects.\,5 and 6 so as to achieve a quantitatively reliable, QCD-connected unification of the pion's valence-quark distribution function (PDF) with, \emph{inter alia}, its distribution amplitudes and elastic electromagnetic form factor.  Whilst this cannot change the essential features of the valence-quark PDF, it will produce some quantitative modifications and one would also thereby obtain predictions for the sea-quark and gluon distributions, which are poorly constrained by existing experiment and theory.  Completing such a picture is crucial as hadron physics enters an era of new-generation experimental facilities, whereat measurements could be made that would better constrain all the pion's parton distribution functions, using techniques such as those discussed in Refs.\,\cite{Holtmann:1994rs,Londergan:1995wp,Londergan:1996vh,Holt:2000cv,Petrov:2011pg}.
The attendant possibilities also provide strong motivation for generalising both our algebraic framework and more sophisticated treatments in order to compute the kaon's PDF.

%

\medskip

\noindent\textbf{Acknowledgments}.
We are grateful for astute remarks by I.\,C.~Clo\"et, B.~El-Bennich, G.~Krein, J.~Segovia and A.\,W.~Thomas; and for the chance to participate in the workshops ``Many Manifestations of Nonperturbative QCD under the Southern Cross'', Ubatuba, S\~ao Paulo, and (CDR, PCT) the ``2$^{\rm nd}$ Workshop on Perspectives in Nonperturbative QCD'', IFT-UNESP, S\~ao Paulo, during which much of this work was completed.
Research supported by:
University of Adelaide and Australian Research Council through grant no.~FL0992247;
Commissariat \`a l'Energie Atomique;
Joint Research Activity ``Study of Strongly Interacting Matter'' (Grant Agreement no.\,283286, HadronPhysics3) under the Seventh Framework Programme of the European Community;
GDR 3034 PH-QCD ``Chromodynamique Quantique et Physique des Hadrons'';
ANR-12-MONU-0008-01 ``PARTONS'';
Spanish ministry Research Project FPA2009-23781;
Department of Energy, Office of Nuclear Physics, contract no.~DE-AC02-06CH11357;
and
National Science Foundation, grant no.\ NSF-PHY1206187.

\vspace*{-2ex}




%




\end{document}